\documentclass[pre,twocolumn,showpacs,amsmath,amssymb,floatfix]{revtex4}
\usepackage{epsfig}
\begin{document}

\title{Dynamical arrest and replica symmetry breaking in attractive colloids}

\author{
A. Velenich$^1$, A. Parola$^2$ and L. Reatto$^3$}

\affiliation{$^1$ Physics Department, Boston University, Boston, MA, USA \\
$^2$ Dipartimento di Fisica e Matematica, Universit\`a dell'Insubria, Como, Italy \\
$^3$ Dipartimento di Fisica, Universit\`a di Milano, Milano, Italy }

\date{\today}

\begin{abstract}
Within the Replica Symmetry Breaking (RSB) framework developed by M.M\'ezard and G.Parisi\cite{MP96} we
investigate the occurrence of structural glass transitions in a model of fluid characterized by 
hard sphere repulsion together with short range attraction. This model is appropriate for the description 
of a class of colloidal suspensions. The transition line in the density-temperature plane
displays a reentrant behavior, in agreement with Mode Coupling 
Theory (MCT), a dynamical approach based on the Mori-Zwanzig formalism. Quantitative differences are 
however found, together with the absence of the predicted glass-glass transition at high density. 
We also perform a systematic study of the pure hard sphere fluid
in order to ascertain the accuracy of the adopted method and the convergence of the numerical procedure. 
\end{abstract}


\maketitle


\section{Introduction}

Understanding the physical conditions leading to dynamical arrest in colloidal suspensions 
and protein solutions is still attracting considerable interest \cite{exp}. 
Colloidal systems are often modeled as simple fluids of (large) 
spherical particles interacting via the hard core repulsion plus, possibly, a short range attraction. The latter 
contribution may have diverse physical origins, according to the specific system considered, like 
the depletion mechanism or steric effects. 

This simple model has been investigated by simulation methods \cite{sim} and liquid state theories 
\cite{liq}. The key dimensionless parameter 
governing the physics of the system is the ratio between the range of attraction ($\Delta$) 
and the hard sphere diameter 
($d$) which, in the following will be taken as the unit of lengths. The equilibrium phase diagram is now well 
known in a wide range of $\Delta$ and the stability of the fluid phase as a function of $\Delta$ 
has been extensively discussed in the literature. 
If the range $\Delta$ is such that at high densities many particles lie within the 
respective ranges of attraction, the overall effect of the attractions is expected \cite{F02} 
to be just a shift in the ground state energy of the system, while the dynamical and thermodynamical 
properties are mostly determined by the hard cores of the particles.
Conversely, if the range of the attractive well is short enough, small displacements of the 
particles can break one or more ``energetic bonds'' and cause significant variations in the energy 
of the microscopic state.
While the hard sphere fluid is a purely entropic system, in this case we have both energetic 
and entropic contribution to the free energy whose relative importance can be tuned by 
varying the temperature and/or the density.

The possible occurrence of dynamical arrest in this model 
has been studied via Mode Coupling Theory (MCT) which indeed 
predicts the existence of two distinct glassy states, called ``attractive'' and ``repulsive'' glass, 
separated in the phase diagram by a first-order transition line terminating in a critical point \cite{D00,Pham}.
This finding might have relevant consequences on the experimentally accessible systems and it is 
therefore important to investigate the problem by alternative techniques. 

The occurrence of a glassy states in statistical physics was 
studied mostly by use of the replica approach in the framework of spin glasses \cite{MPV87}. 
When considering a system with quenched disorder, it is natural to introduce the whole ensemble of possible realizations of the quenched variables. For a variety of models it turns out that the values of many quantities of physical interest equal, in the thermodynamic limit, their average over the quenched variables. Subsequent studies revealed that in order to produce glassy behavior, an externally imposed quenched disorder is not essential 
because the frustration necessary for slow dynamics and dynamical arrest can be self-generated by inter-particle
interaction.

Throughout the formalism, different replicas are in principle indistinguishable and the equations for a replicated system are symmetric with respect to permutations of replicas; yet, replica symmetry breaking (RSB) solutions have proved to be the appropriate solutions for the low-temperature phases of several models. For such solutions, the correlation functions between different replicas do depend on the replica indices.

A general framework often used to account for the slow dynamics of glasses is based on the analysis of 
a free energy $F$, considered as a functional of the \emph{averaged} density profile\cite{note1}.
The key feature is that at low temperatures $F$ is expected to develop a multi-minima structure and, if the free energy barriers are high enough with respect to $k_BT$, the configuration space becomes essentially disconnected, leading to a non-ergodic behavior.
The basic assumption underlying the mean-field approach presented in this paper is the modeling of the glassy non-equilibrium dynamics through such a free energy landscape. Hence, since the whole formalism of equilibrium thermodynamics promptly applies, we can refer to a glassy state as a proper \emph{phase} of the system. The dynamic crossover observed experimentally is correspondingly replaced by a sharply defined phase transition.

In this context, the introduction of replicas is a procedure intended to unveil a possible multi-minima structure of the free energy: it is natural to expect that for two replicas in the same local minimum of $F$ the correlation is larger than for replicas in different minima. Hence, the appearance of a non-trivial pattern of correlation among replicas (i.e. the breaking of the replica symmetry) is the signature of the complex free energy landscape characteristic of a glassy phase. 
In the following, by studying the correlation functions between different replicas of a fluid system, we confirm the existence of a RSB phase transition. The comparison of the resulting phase diagram with experimental and numerical data on dynamical arrest in systems with short range attractive interactions strongly supports the interpretation of such a transition as an ideal glass transition.

We point out that by ``ideal'' glass transition it is usually meant a transition defined by a thermodynamic singularity \cite{PZ05}. With the replica method, \emph{two} critical densities are obtained: the lower density signals the appearance of a complex free energy landscape\cite{note2}
with no thermodynamic singularity. At the higher density the complexity vanishes non-smoothly and this gives a true thermodynamic transition. In \cite{MP96} it was shown that the replica approach we are going to use is not suitable for the description of the glassy phase and leads to incorrect results for the thermodynamic transition. For the time being we do not exclude that a more accurate numerical study of the thermodynamic transition might yield different results (see table in section \ref{HS} and consider that the grid used in \cite{MP96} had only $N=128$ points), however in this paper we limit our attention to the first transition, commonly known as ``dynamical'' glass transition. 

We first review the basic equations derived in Ref. \cite{MP96} then we perform a careful study of the hard sphere system, in order to test the accuracy of the method and to estimate the numerical uncertainty of our results. 
Finally, we investigate the effects of a short range interaction and compare our results with MCT predictions. 
We anticipate that, although the transition line we find is rather similar to those of MCT clearly showing the
predicted reentrant behavior, we have no evidence 
in favor of glass-glass transition, within the adopted approximations.  
  

\section{The Replica Method for structural glasses}

According to the previous discussion, in order to implement the replica symmetry
breaking scheme for a model of fluid, we need a formal expression for the free energy 
functional of a mixture of identical copies of the original system. 
In the following, the indices $a$ and $b$ identify different replicas. Such a generalized 
free energy functional $F$ depends on the two body interactions among particles $U_{ab}(r)$
(both intra and inter-replicas) and on the two point correlation functions $g_{ab}(r)=h_{ab}(r)+1$, whose entries are labeled by replica indices: ordinary (intra-replica) 2-point functions on the diagonal and inter-replica correlation functions as off-diagonal elements.
The original physical system can be recovered by setting $n=1$. 
The key property of the 
functional $F$ is to attain its global minimum when the pair correlations $g_{ab}(r)$ assume the
physical value corresponding to the given interaction $U_{ab}(r)$. RSB does occur when a non trivial
minimum (i.e. a solution with finite correlations among different replicas) 
is present in the physical limit of uncoupled replicas $U_{ab}(r)=U(r)\delta_{ab}$. 
For a homogeneous system, the general structure of the free energy density functional is
\begin{equation}
F[U,g] = \sum_{a,b}\,\frac{\rho_a\rho_b}{2}\,\int \mathrm{d}^3\mathbf{x} \, g_{ab}(x)U_{ba}(x) - T\,S[g]
\label{free}
\end{equation}
The internal energy contribution is exactly given by the first term of (\ref{free}) while
the entropy density $S$ is known to be a functional of the pair correlation functions alone, which can be 
written \cite{green} as a sum of $p$-particle contributions: $S=\sum_p s_p$.
By use of the Gibbs-Bogoliubov inequality \cite{HMD06} it is easy to prove that the functional $F$ defined 
in (\ref{free}) satisfies the requirements previously mentioned. Unfortunately, a closed expression
for the entropic contribution $S$ is not available and we have to resort to some approximation. 
Following Ref. \cite{MP96} we keep the exact two body term of the excess entropy $s_2$ 
\begin{equation} \label{s2}
s_2 = - k_B\,\sum_{a,b}^{n}\,\frac{\rho_a\rho_b}{2} \int \mathrm{d}^3\mathbf{x}\, 
[\, g_{ab}(x) \ln g_{ab}(x) - h_{ab}(x) ]
\end{equation}
while approximating the
residual contribution $\Delta S = S-s_2$ in terms of the pair distribution function:
\begin{equation} \label{deltas}
\Delta S = - k_B\,\frac{1}{2} \int \frac{\mathrm{d}^3\mathbf{q}}{(2\pi)^3} \mathrm{Tr} L(\rho\mathbf{h}(q))
\end{equation}
where the function $L(y)$ is defined as
\begin{displaymath}
L(y)=-\ln(1+y)+y-\frac{y^2}{2}
\end{displaymath}
and $\mathbf{h}(q)$ represents the $n\times n$ matrix whose elements are the Fourier transforms of the correlation functions $h_{ab}(r)$.
This form follows form the resummation of the infinite class of non crossing diagrams in the
formal expansion of the entropy as a functional of $h(r)$. 
Collecting all terms together we find the final expression for the free energy functional
which, in the physical limit of uncoupled replicas becomes:
\begin{eqnarray} \label{repF}
2\beta F &=& \rho^2 \int \mathrm{d}^3\mathbf{x} \sum_{a,b}^{n} \Big( g_{ab}(x) 
[\ln g_{ab}(x) + \beta U(x)\delta_{ab}] \nonumber \\
&-& h_{ab}(x) \Big) + \int \frac{\mathrm{d}^3\mathbf{q}}{(2\pi)^3} \mathrm{Tr} L(\rho\mathbf{h}(q))
\end{eqnarray}

In the case of a single component ($n=1$), the minimization of the free 
energy functional (\ref{repF}) leads to the usual HNC integral equation for the pair correlation:
\begin{eqnarray} \label{HNC} 
\ln g(x) &=& -\beta U(x) + W(x) \nonumber \\
W(x) &=& \int \frac{\mathrm{d}^3\mathbf{p}}{(2\pi)^3} e^{-i\mathbf{p \cdot x}} \frac{\rho h^2(p)}{1+\rho h(p)}
\end{eqnarray}
Yet, by keeping $n$ unspecified, a second equation is obtained: the appearance of multiple solutions for this new equation signals a phase transition. 
Given the replicated free energy, the fundamental ansatz is a 1-step RSB. As we said, in general, a RSB directly mirrors a change in the ``effective topology'' of the configuration space which takes on a multi-valley structure, with ``disconnected'' single-valley domains. By choosing a 1-step RSB we impose only two possible patterns for the functional form of the correlations between replicas: either two replicas are in the same minimum of the free energy or they are in different minima; in the latter case we \emph{assume} that the correlation of every pair of replicas is the same, no matter what the two minima are. In particular, for a structurally disordered material, we expect that any spatial correlation between two replicas in different minima should vanish: this conjecture
follows from the observation that the correlation between different disordered configurations, despite possible local similarities, should vanish in the thermodynamic limit, when averaged over the infinite volume of the system. A formal argument for the relevance of the 1-step RSB in structural glasses and a physical interpretation of $m$ (defined below) in terms of effective temperature is presented in \cite{M95}.
\begin{figure} \label{RSBpattern}
\begin{center}
\epsfig{figure=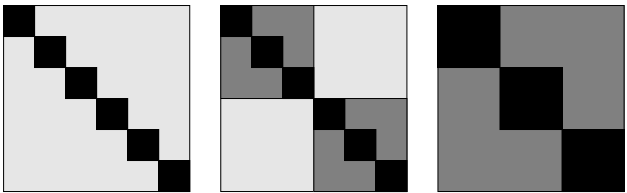,height=2.5cm}
\caption{
Replica symmetric (left) and 1-step replica symmetry breaking (center) form of the correlation matrix for a replicated system. The black squares represent single elements ($g_*$), while the dimension of the $n \times n$ matrix (light-gray for $g_0$) and of the $m \times m$ blocks (dark-gray for $g_1$) is, in general, not fixed. In the figure, $n=6$ and $m=3$. Since in the physical case of zero replica coupling the correlators corresponding to the light gray area vanish, the numerical computations can be simplified by considering only one of the blocks (right).
}
\end{center}
\end{figure}
In terms of correlation matrices, the standard procedure is to group the $n$ replicas into $n/m$ sets with $m$ replicas each. Ideally, the $m$ replicas within a block are in the same minimum of $F$. The ``topological" structure of the matrix of correlation functions $g_{ab}$ is pictorially depicted in Fig. 1: The black squares represent the ordinary two-point correlation functions for a fluid ($g_*$); the correlation functions in the light-gray blocks ($g_0$) are assumed to be unity (different minima are unrelated) and the correlation functions in the dark-gray blocks ($g_1$) may be non trivial (replicas in the same minimum). 
Since each term in (\ref{repF}) has a block-diagonal form and $h_0=0$, the free energy \emph{per replica} turns out to be independent of $n$ and the problem reduces to the study of any of the $n/m$ blocks
of dimension $m \times m$ \cite{note3} (fig.\ref{RSBpattern}c). Expressing $F$ in terms of $g_*$ and $g_1$:
\begin{widetext}
\begin{eqnarray}
\label{FRSB}
\frac{2\beta F}{n \rho^2} &=& \int \mathrm{d}^3\mathbf{x} \Big( g_*(x) [\ln g_*(x) + \beta U(x)]-h_*(x) - 
 (1-m)[g_1(x)\ln g_1(x) - h_1(x)] \Big) 
- \int \frac{\mathrm{d}^3\mathbf{q}}{(2\pi)^3} \Big( \frac{1}{m\rho^2} \ln[1+ \nonumber \\
&+& \rho h_*(q) -(1-m)\rho h_1(q)] - \frac{h_*(q)}{\rho} + \frac{h_*^2(q)}{2} 
-\frac{1-m}{m\rho^2}\ln[1+\rho h_*(q)-\rho h_1(q)] - (1-m)\frac{h_1^2(q)}{2} \Big)
\end{eqnarray}
\end{widetext}
The stationary condition with respect to $g_*$ is the usual HNC equation (\ref{HNC}), while,
due to a simplification of factors $(1-m)$, the extremal equation for the inter-replica correlation function $g_1$ is non-trivial also in the ``physical'' case $m\to 1$:
\begin{eqnarray} \label{h1}
\ln g_1(x) &=& W_1(x) \nonumber\\
W_1(q) &=& \frac{\rho h_*^2(q)}{1+\rho h_*(q)}
- \frac{\rho[h_*(q)-h_1(q)]^2}{1+\rho[h_*(q)-h_1(q)]}
\end{eqnarray}
Because of the non linear structure of equation (\ref{h1}), we can now hope to obtain non-trivial solutions $g_1(x)$ which would otherwise be ruled out by the replica-symmetry of the partition function.
Indeed, we will show that the extremal equations for the free energy (\ref{FRSB}) admit non-trivial correlations between replicas (i.e. a non trivial $h_1(x)$), even for \emph{vanishing} inter-replica coupling. 
The typical behavior of such a solution is shown in figure 2.
\begin{figure} \label{figgr}
\centering
\epsfig{figure=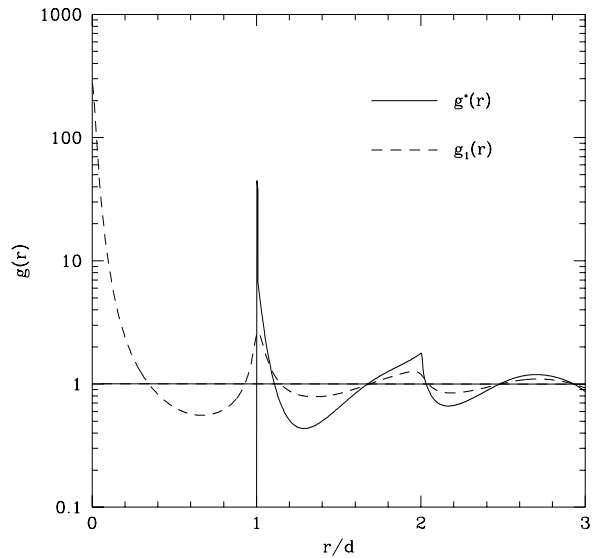,width=8cm}
\caption{An example of correlation functions for sticky spheres close to the dynamical glass transition.
The thermodynamic state is $\rho=1.1878$ $T=0.6$ and the data refer to the system with $\Delta =1/64$.}
\end{figure}

To summarize, starting with a one component model of fluid, by introducing $m$ identical copies and taking the limit $m\to 1$, after simple algebraic manipulations of the approximate free energy functional (\ref{repF}), we obtain the usual HNC equation for the two-point function \emph{and} one additional equation which describes correlations between different replicas. Since for $m=1$ an inter-replica correlation has apparently no meaning, we could na\"\i vely interpret $h_1(x)$ as an infinite time limit of a self-correlation function: The presence of a non trivial solution $h_1(x)$ points to ergodicity breaking. This point is further discussed in section \ref{ggtrans}.


\subsection{Solution of the equations} \label{solution}

For any density, $h_1(x) \equiv 0$ is a solution of equation (\ref{h1}). By linearizing (\ref{h1}) around $h_1(x) \equiv 0$ it can be shown that, if another solution exists, it does not bifurcate from the trivial one. However, if we introduce a strong enough \cite{TDF00} coupling between replicas, the linearized equation does not forbid a continuous bifurcation of solutions and a second solution has indeed been found. 
When the coupling is eventually turned off, one of these solutions converges to $h_1(x)\equiv 0$ (``liquid'' solution), while the other one maintains a spatial structure (``glassy'' solution). This signals a multi-minima free energy, which we interpret as a non-ergodic phase. In fact, the free energy is now a functional of $g_*$ and $g_1$: the occurrence of a second minimum with $g_1 \neq 0$, mirrors the appearance of many minima in $F[\langle \rho(x) \rangle]$, \cite{CFP99}. The glassy solution of equation (\ref{h1}) does not exist for all values of the 
control parameters but only  in a region of the density-temperature plane, 
and the boundary between the 1-solution and the 2-solutions domain in the phase diagram marks 
the glass transition.

The two-point functions $h_*(x)$ and $h_1(x)$ are computed by the numerical solution of equation (\ref{h1}). 
We first introduce a mesh in the radial coordinate, thereby reducing the unknown function $h_1(x)$ to a 
set of $N$ discrete values ($h_1(x) \mapsto h_1(x_i) \quad i=1,..,N$). Then we use the iterative 
Newton-Raphson method for solving the resulting set of non linear equations . 
In order to find a non-trivial result, we proceed as follows: starting from the ``liquid'' 
solution at low density ($\rho < \rho_t$), we introduce a finite attractive coupling $\varepsilon$ among replicas
\cite{Picard} and \emph{subsequently} increase gradually the density until the continuous bifurcation 
takes place. We then follow the glassy solution and finally reduce to 0 the coupling between replicas, 
obtaining a non-trivial correlation function $h_1(x)$ even for vanishing inter-replica coupling.
The transition density $\rho_t$ can then be pinpointed by decreasing (at zero inter-replica coupling) the density until the glassy minimum of the free energy disappears. The values we report in the following for the transition densities are the first ones for which our algorithm does not find a solution.
Theoretically, the introduction of a coupling among replicas provides a way to discover multiple degenerate states, in this case it is motivated only by the computational convenience to obtain a \emph{continuous} bifurcation of the new minimum of the free energy from the known one. 


\section{The hard sphere fluid} \label{HS}

The glass transition in the hard sphere fluid has already been studied by the replica method in Refs. \cite{MP96} and \cite{TDF00}. Here we consider the dependence of the critical density on the choice 
of the discretization parameters used in the a numerical solution, and we provide an estimate 
of the transition density $\rho_t$.

Given the number of points $N$ for the discretization of the correlation functions, the only arbitrary choice is the mesh size $a=1/N_d$ (where $N_d$ is the number of points used to describe a hard sphere diameter). It then follows that we can describe a correlation function $h(r)$ only for $r \in [0,N/N_d]$.
The physical limit should require both $a \rightarrow 0$ and $N/N_d \rightarrow \infty$. 

After choosing a value for $N_d$, we proceed by calculating the transition density for grids with more and more points $N$ while keeping $N_d$ fixed: for $N$ large enough, the value of the transition density becomes independent of $N$ and we can label it as $\rho_t(N_d)$.
Repeating the procedure for larger and larger $N_d$'s the sequence of values we obtain converges to a finite value. 
Our results for the hard sphere fluid are summarized in table \ref{tab1}.

\begin{table}
\begin{tabular}{|c|c|c|c|c|c|c|c|c|c|} 
\hline 
 & &$N$& 128 &	256 &	512 &	1024 &	2048 &	4096 &	8192 \\
\hline
$\alpha$& $N_d$& \multicolumn{8}{|c|}{} \\
\hline
5&\multicolumn{2}{|l|}{32} &   1.1124&	1.1031&	1.1026&	1.1026&	1.1026&	1.1026&	$\mathbf{1.1026}$\\
\hline     
6&\multicolumn{2}{|l|}{64} &   	1.1364&	1.1518&	1.1391&	1.1396&	1.1396&	1.1396&	$\mathbf{1.1396}$\\
\hline
7&\multicolumn{2}{|l|}{128} & 	  -&	1.1497&	1.1681&	1.1552&	1.1560&	1.1559&	$\mathbf{1.1560}$\\
\hline
8&\multicolumn{2}{|l|}{256} & 	  -&	  -&	1.1553&	1.1753&	1.1627&	1.1636&	$\mathbf{1.1636}$\\
\hline
9&\multicolumn{2}{|l|}{512} & 	  -&	  -&	  -&	1.1578&	1.1787&	1.1663&	$\mathbf{1.1672}$\\
\hline
10&\multicolumn{2}{|l|}{1024}& 	  -&	  -&	  -&	  -&	1.1590&	1.1804&	$\mathbf{1.1680}$\\
\hline
11&\multicolumn{2}{|l|}{2048}& 	  -&	  -&	  -&	  -&	  -&	1.1595&	1.1812\\
\hline
12&\multicolumn{2}{|l|}{4096}& 	  -&	  -&	  -&	  -&	  -&	  -&	1.1598\\
\hline
\end{tabular}
\caption{Glass transition density $\rho_t$ for different choices of the discretization parameters: 
total number of mesh points $N$ and number of points inside the hard sphere diameter $N_d$.}
\label{tab1}
\end{table}
It is clear that only when $N/N_d \gtrsim 4$ the values $\rho_t(N_d)$ are almost independent of $N$.
If we now consider the values obtained for the 8192-point grid, by \emph{arbitrarily} interpolating the values $\rho_t(N_d)$ with a function of the form:
\begin{equation} \label{intrp}
\rho_t(N_d \equiv 2^{\alpha})=\rho_t - A \exp(-k\alpha)
\end{equation}
we can extrapolate a limiting value $\rho_t$.

\begin{figure}
\centering
\epsfig{figure=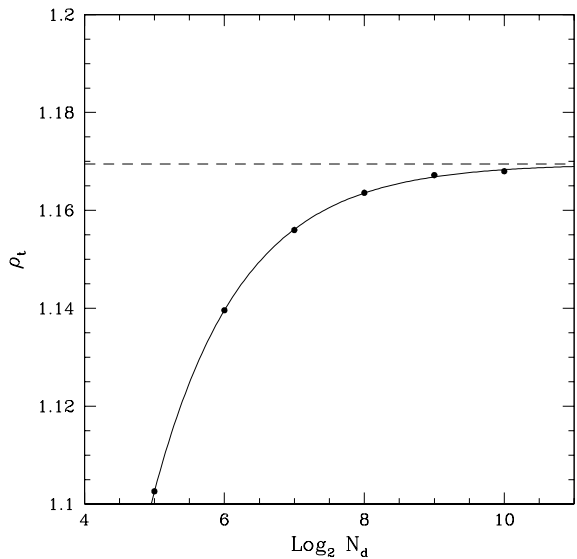,width=8cm}
\caption{Dependence of the critical density on the grid: the dots represent the data in the last column of the table, the dashed line is the best fit using equation (\ref{intrp}).}
\end{figure}

Depending on the inclusion of the first 5 or 6 points in the interpolation ($\alpha = 5,..,9$ or $\alpha=5,..,10$ respectively, the boldface entries in the table), we obtain $\rho_t=1.1699$ or $\rho_t=1.1695$ (corresponding to a packing fraction $\eta_t \approx 0.612$); this should be considered the final prediction for the transition density of a hard sphere fluid in HNC approximation.
This result is rather close to the random close packing limit ($\eta_{rcp}\sim 0.64$) and is considerably higher 
than the prediction of MCT ($\eta_t\sim 0.516$) \cite{D00}. We note that recent experiments suggest that, in the absence of 
gravity, the glass transition in ``hard sphere colloids" is in fact remarkably close to $\eta_{rcp}$ \cite{kegel}.  

For the sake of completeness we obtained a similar table for the thermodynamic glass transition also (up to a 4096-point grid): the extrapolation of the asymptotic density yields $\rho_g \simeq 1.189$, which is equivalent to a packing fraction $\eta_g \simeq 0.623$.


\section{The Square Well fluid}

In order to model a colloidal suspension, we now investigate the effects of the presence of a
short range attraction by adding an attractive square well to the repulsive hard core. The interaction potential is thus:

\begin{equation} \label{SW}
U(r) = \left\{
\begin{array}{ll}
\infty & r \leq d \\
-U_0 & d < r \leq d+\Delta \\
0 & d+\Delta < r \\ 
\end{array} \right.
\end{equation}
As usual, we take $d$ as the unit of lengths.
The application of the replica method is straightforward, but the phenomenology is now potentially richer, since both the density and the temperature are meaningful thermodynamic parameters.  
By inserting the potential (\ref{SW}) into Eq. (\ref{HNC}) and applying the procedure described in section \ref{solution} for different temperatures, we determine a whole transition line $\rho_t(T)$ in the $\rho$-T plane. 

In figure 4 we summarize the numerical study of the short range limit of the square well fluid. The results agree with the qualitative picture following from the above considerations: at high temperatures the glass transition line in the $\rho$-T plane is almost vertical and, as for the hard sphere fluid, and the glass transition is governed by excluded volume effects. At low temperatures, the transition line displays a strong dependence on the temperature and only a slight dependence on the density: the dynamical arrest is driven mainly by the stickiness of the spheres. The replica method correctly captures both regimes, extending the transition line to densities well below $\rho_t$.
\begin{figure} \label{phdiagr}
\centering
\epsfig{figure=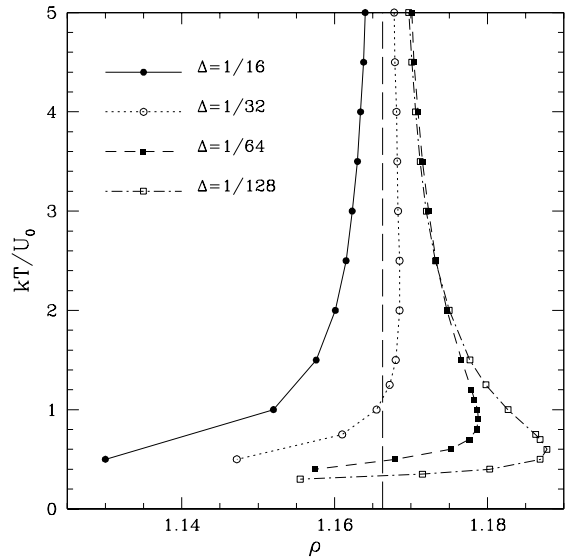,width=8cm}
\caption{Phase diagram for the square well fluid for different well widths $\Delta$. The dashed line represents the hard sphere transition density computed with the same discretization parameters.}
\end{figure}
Moreover, in a region of the phase diagram, the ``competition'' between energy and entropy stabilizes the liquid phase at densities \emph{larger} than $\rho_t$. This phenomenon, commonly referred to as ``reentrant behavior'', is also reproduced.

The computations were performed on a 4096-point grid (512 points for a hard sphere diameter). Because of the very short range of the attraction, few points were available to describe the narrowest wells. Due to the non linearity of the integral equations, an estimate of the errors introduced by our discretization is quite difficult; yet, the \emph{qualitative} behavior of the transition line seems to be well established.
As an empirical test of the reliability of our results for wells described with very few (2 or 4) points on the grid, we repeated the computations for part of the phase diagram in figure 4 using a looser grid (2048-point grid, 256 points for a hard sphere diameter and a corresponding rescaling by a factor of 2 for the attractive wells). The numerical results agreed within 0.2\%.

The convergence of our algorithm becomes rather delicate as we follow the liquid-glass transition line to low-temperatures/low-densities and we defer to future studies the behavior of the transition line in such a limit. An intrinsic limitation of our approach, comes from the equilibrium character of the replica method, which cannot be applied below the spinodal decomposition line.

As already found in the pure hard sphere fluid, the transition densities are
significantly higher ($\sim 15\%$) than those obtained by MCT \cite{D00} and the corresponding 
critical temperatures are lower, roughly by a factor $2$. 
The reentrant behavior of the glass transition, found by MCT, is confirmed by RSB method when $\Delta$ is sufficiently small. 
However, while in MCT the reentrance appears 
for $\Delta \lesssim 5\%$, in our calculations a similar shape is fond only for $\Delta \lesssim 1.5\%$. 

\subsection{The glass-glass transition}
\label{ggtrans}

An interesting phenomenon that might be within the scope of the replica approach is the glass-glass transition predicted by MCT. In particular, in a dynamical theory, the localization length  of a particle can be defined: $\sigma = \lim_{t\rightarrow\infty} \langle |\mathbf{r}(t)-\mathbf{r}(0)|^2 \rangle$. In a solid or in a glass, $\sigma$ attains a finite value, while it diverges in the fluid phases. It has been shown \cite{D00} that $\sigma$ undergoes a discontinuous change across the glass-glass transition line, decreasing from its value in the repulsive glass to a smaller value comparable to the attractive well width, in the attraction induced glass.

In our formalism, this information is not directly accessible. However, a hint on the local structure 
of the glassy state can be obtained by defining the quantity  
$\mathrm{P}(r)=4\pi r^2 g_1(r)$ which represents the probability (per unit length) 
that two particles belonging to different replicas lie at distance $r$ one another 
and can be thought of as the infinite-time limit of a dynamical self-correlation function \cite{FP95}. 
In figure 5  we plot such a probability for values of the control parameters which, according to the MCT, should identify distinct glassy phases. In all cases, the probability distribution is uniformly spread over a region of the order of a hard sphere volume. Even considering the local maximum of $\mathrm{P}(r)$ for $r \simeq 0.05$ (see figure 5), comparable to MCT localization length $\sigma$, its position varies \emph{smoothly} with the control parameters and its value is always close to the one obtained for the pure hard sphere system.
\begin{figure}\label{fig5}
\centering
\epsfig{figure=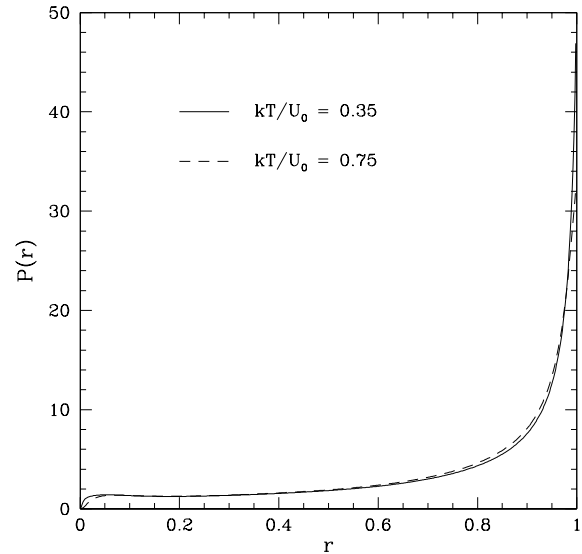,width=8cm}
\caption{$\mathrm{P}(r)$ for the same density ($\rho = 1.19$) and $\Delta=1/128$ but different temperatures.} 
\end{figure}
\begin{figure}\label{fig6}
\centering
\epsfig{figure=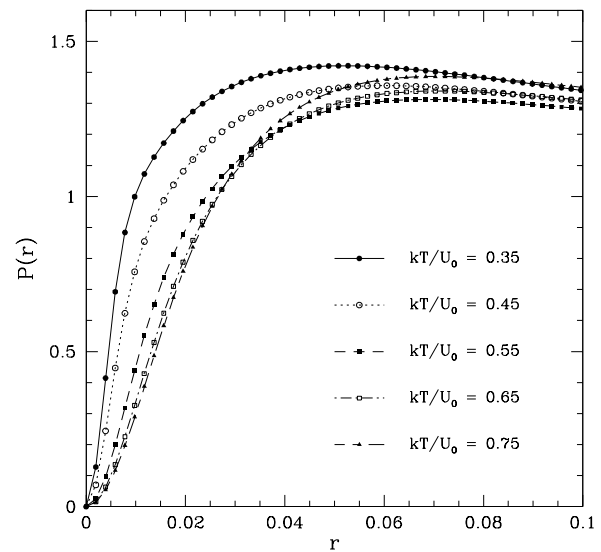,width=8cm}
\caption{Detailed structure of $P(r)$ at short distance for the data of Fig. 5. As the temperature decreases at constant density $\rho = 1.19$ (see the vertical cut in figure 4) no singular behavior of $P(r)$ is observed.}
\end{figure}
We cannot exclude that a different choice for the replica symmetry breaking scheme (e.g.~a 2-step RSB) might lead to a well-defined glass-glass transition: the 2-step RSB would then represent the splitting of the free energy minima revealed with the 1-step RSB into sub-minima representing a further trapping of the caged particles within the range of the attractive wells. This possibility is currently under investigation. 


\section{Conclusions}

In this paper we first investigated the occurrence of RSB in a model of hard spheres, 
providing an accurate finite size scaling for the transition density.
The mean-field replica approach implies a sharp (dynamical) glass transition; the asymptotic result we obtained for hard spheres, at a packing fraction of $\eta_{Repl} \simeq 0.612$ should be compared with the MCT value $\eta_{MCT} \sim 0.52$ and the commonly accepted value $\eta \simeq 0.58$ for the slowing down of the dynamics of colloidal suspension and simulated hard sphere systems 
(even though experiments conducted in micro-gravity conditions 
\cite{kegel} have recently raised this value, significantly closer to the random close packing limit).
By introducing attractive interactions we obtained the glass transition line in the density-temperature
plane. The expected reentrant behavior related to the change between ``repulsive" and ``attractive" glass is
reproduced within the RSB approach. Quantitative discrepancies with respect to MCF are found for the square well fluid, 
analogously to the hard sphere case. Moreover, contrary to MCT, the attractive and repulsive glassy regimes are 
smoothly connected with no sign of a sharp transition. These results have been obtained within a 
simple approximation of the more general RSB method: we just considered the {\sl one step} RSB scheme and
we evaluated the entropy functional by a HNC-like expression. It is clearly desirable to go beyond 
these limitations in order to check the qualitative and quantitative stability of the results obtained
in this work. 


\section{Acknowledgments}

We warmly thank K.A.Dawson, F.Thalmann and C.Dasgupta for valuable discussions and suggestions. This work is supported 
by MCRTN-CT-2003-504712. 



\begin{thebibliography}{99}
\bibitem{note1} Starting with a position dependent pinning field $\mu(x)$ for the density, the 
appropriate thermodynamic potential can be computed and applying a Legendre transform 
$F[\langle \rho(x) \rangle ]$ is then obtained.
\bibitem{note2} The number of minima $\mathcal{N}$ is usually written in terms of a ``complexity'' 
$\Sigma (\rho , T)$ which represents the contribution to the entropy deriving from the presence 
of several minima: $\mathcal{N}=e^{N\Sigma}$, where $N$ is the number of particles in the system.
\bibitem{note3} We note that within each minimum of the free energy (within each $m \times m$ block), 
the replica symmetry of the correlation matrix is restored.
\bibitem{Picard} For $U_{ab}$ we used a gaussian potential with $\sigma=1/4$.
The coupling should be strong enough to obtain a continuous bifurcation of solutions: 
an example of ``extended'' phase diagram $\rho - \varepsilon$ is shown in Ref. \cite{TDF00}. 
But since we are interested only in the limit $\varepsilon\rightarrow 0$, there is no need to allow for 
$h_0(x) \neq 0$ as in \cite{TDF00}, and the form of the coupling can be chosen as block-diagonal, 
with no coupling between replicas in different blocks. This makes the dependence on $n$ trivial and 
reduces by a factor two the computing time, since only $h_1(x)$ and not $h_0(x)$ should be calculated 
(The computation of $h_*(x)$ is performed through a Picard method which is considerably faster than 
the minimization procedure necessary for $h_1(x)$).
\end{thebibliography}
\end{document}